\documentclass[10pt,amsmath,amssymb,aps,prl,twocolumn]{revtex4-2}

\usepackage{graphicx} 

\usepackage{bm}
\usepackage{color}
\usepackage{siunitx}

\usepackage{blindtext}

\usepackage{hyperref}

\newcommand{\beq}{\begin{equation}}
\newcommand{\eeq}{\end{equation}}
\newcommand{\bea}{\begin{eqnarray}}
\newcommand{\eea}{\end{eqnarray}}
\newcommand{\bec}{\begin{center}}
\newcommand{\enc}{\end{center}}
\newcommand{\bfr}{\begin{flushright}}
\newcommand{\efr}{\end{flushright}}
\newcommand{\abs}[1]{\left|{#1}\right|}

\begin{document}

\title{On-demand generation and characterization of a microwave time-bin qubit}

\author{J. Ilves$^{1*}$}
\author{S. Kono$^{1}$}
\author{Y. Sunada$^{1}$}
\author{S. Yamazaki$^{1}$}
\author{M. Kim$^{1}$}
\author{K. Koshino$^{2}$}
\author{Y. Nakamura$^{1,3*}$}%

\affiliation{$^{1}$Research Center for Advanced Science and Technology (RCAST), The University of Tokyo, Meguro-ku, Tokyo 153-8904, Japan}
\affiliation{$^{2}$College of Liberal Arts and Sciences, Tokyo Medical and Dental University, Ichikawa, Chiba 272-0827, Japan}
\affiliation{$^{3}$Center for Emergent Matter Science (CEMS), RIKEN, Wako, Saitama 351-0198, Japan}

\date{December 6, 2019}

\begin{abstract}
Superconducting circuits offer a scalable platform for the construction of large-scale quantum networks where information can be encoded in multiple temporal modes of propagating microwaves. Characterization of such microwave signals with a method extendable to an arbitrary number of temporal modes with a single detector and demonstration of their phase-robust nature are of great interest. Here we show the on-demand generation and Wigner tomography of a microwave time-bin qubit with superconducting circuit quantum electrodynamics architecture. We perform the tomography with a single heterodyne detector by dynamically changing the measurement quadrature with a phase-sensitive amplifier independently for the two temporal modes. By generating and measuring the qubits with hardware lacking a shared phase reference, we demonstrate conservation of phase information in each time-bin qubit generated.
\end{abstract}


\maketitle

In the past few decades, quantum bits implemented as superconducting circuits have become promising candidates for building blocks of large-scale quantum computers~\citep{bib:krantz2019quantum, bib:wendin2017quantum, bib:gambetta2017building, bib:gu2017microwave}. To increase the scalability of these architectures, robust methods of generating single photons for quantum computation in the propagating modes and for transferring information between multiple superconducting qubits over relatively long distances are of recent interest. In the optical domain, different photonic qubit encodings have been demonstrated before for such purposes~\citep{bib:kok2007linear}. However, optical single-photon generation protocols are often probabilistic rather than deterministic, limiting success probability~\citep{bib:bunching}. Moreover, conversion of quantum information stored in superconducting qubits operated in the microwave regime to optical photons suffers from low efficiency and limited bandwidth~\citep{bib:lambert2019coherent,bib:lauk2019perspectives}. Schemes focused on generating photons at microwave frequencies and their characterization are therefore of great interest.

Photonic qubit encoding can be realized by constructing a set of computational basis states with one or more orthogonal modes of light. In the microwave regime, single-rail (single-mode) encoding has been demonstrated by using the photon number states of a propagating microwave qubit to transfer information between two superconducting qubits over a transmission line with fidelity close to 0.8.~\citep{bib:shapedphotons,bib:kurpiers2018deterministic,bib:devoretbell,bib:schuster2018}.  However, photon loss reduces the transfer fidelity greatly since decayed photon states cannot be distinguished readily. In addition, the phase information in a single-rail photonic qubit state is stored as the relative phase between the propagating qubit mode and a separate phase reference. Thus, the reference must be shared between any hardware operating the nodes of a quantum network that the photonic qubit will interact with, reducing the practicality of single-rail encoding in large networks.

As an alternative to the single-rail encoding, dual-rail (dual-mode) encoding has been demonstrated in the optical regime in the form of polarization~\citep{bib:polar1, bib:polar2, bib:polar3} and time-bin qubits~\citep{bib:opticaltimebin, bib:opticaltimebincomputation}. Occupation of a single photon in one of two orthogonal temporal modes functions as the basis of the time-bin qubit. Time-bin encoding allows one to readily determine loss of information during transfer with a photon number parity measurement~\citep{bib:kok2007linear,bib:parity}, and the qubit state is more robust against dephasing since the phase information is stored in the relative phase between the two temporal modes. Thus, time-bin qubits do not require sharing of a phase reference~\citep{bib:Bartlett2007}. Due to these favorable properties, a linear optical scheme for quantum computation with time-bin qubits has been proposed~\citep{bib:kok2007linear}. However, only the loss-robustness of the microwave time-bin qubit has been demonstrated. The demonstration was based on discrete-variable measurements of superconducting qubits as a part of a transfer protocol, thus being limited to a single qubit of information~\citep{bib:kurpiers2018quantum}. A different approach is necessary for full state tomography of a general two-temporal mode state or cluster states with multiple modes and qubits of information. Ideally, for a scalable characterization method, only a single detector should be necessary regardless of the number of modes.

%
\begin{figure*}[t]
\begin{center}
  \includegraphics[width=166mm]{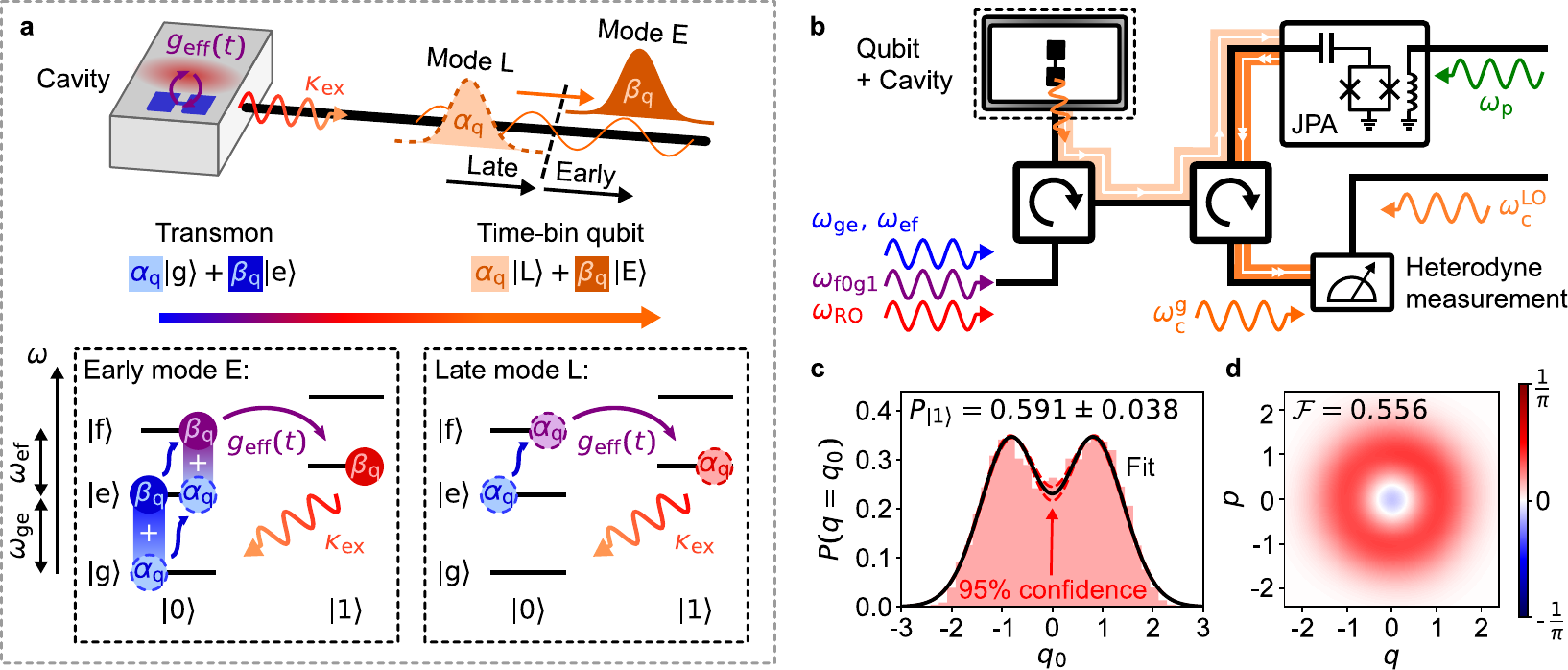} 
\caption{
\textbf{Dynamics of time-bin qubit generation and measurement setup.} (a)~Driven interaction between a superconducting qubit and 3D cavity for the generation of a microwave time-bin qubit propagating along a transmission line. The two energy diagrams for the qubit--cavity system describe the generation protocol.
(b)~Simplified configuration for generating and measuring a time-bin qubit at frequency $\omega_\textrm{c}^\textrm{g}$ with Josephson parametric amplifier (JPA) realized heterodyne measurement. Three different microwave sources are used in the experiment to generate signal at the qubit control frequencies $\omega_\textrm{ge}$ and $\omega_\textrm{ef}$, $|\textrm{f}0\rangle$--$|\textrm{g}1\rangle$ transition frequency $\omega_\textrm{f0g1}$, dispersive cavity readout frequency $\omega_\textrm{RO}$, JPA pump frequency $\omega_\textrm{p}$, and demodulation local oscillator frequency $\omega_\textrm{c}^\textrm{LO}$. (c)~Measured marginal distribution of a single-rail single-photon qubit state (red histogram) as a function of a given quadrature of the generated signal. The black line represents a theoretical fit to the data with 95\% confidence intervals (red dashed line). (d)~Reconstructed Wigner function of the signal with quadratures $q$ and $p$ defined corresponding to $[q, p] = i$. The data in the figures has not been corrected for detection inefficiency.
} 
  \label{fig:f1} 
\end{center}
\end{figure*}
%

In this work, we experimentally demonstrate on-demand generation of microwave time-bin qubits with a superconducting transmon qubit~\citep{bib:transmon} and show how the time-bin qubit retains phase information and can be loss-corrected. Our scheme allows us to generate and shape the single-photon wave packet as well as to generate any superposition state of the time-bin qubit with variable spacing between the temporal modes. We perform Wigner tomography of microwave signal in two temporal modes by measuring the quadrature distributions with a flux-driven Josephson parametric amplifier~\citep{bib:JPA} and a single heterodyne detector~\cite{bib:kono}. With the JPA, we can rapidly change the measurement quadrature for each temporal mode independently in a single shot. We reconstruct the quantum state of the signal with a maximum-likelihood method~\citep{bib:maximumlikelihood,bib:kono}. We compare the state preparation fidelity of the dual-rail time-bin qubit with a single-rail number-basis qubit and a transmon qubit. We demonstrate that correcting photon loss of the time-bin qubit state improves the fidelity significantly. By removing the phase-locking between the single photon source and the detector, we observe that the single-rail photonic qubit state dephases completely due to the lack of a stable phase reference, while the time-bin qubit state is unaffected. This demonstrates that the phase information of the dual-rail qubit is contained in the relative phase between the two modes and that using the time-bin qubit in a quantum network does not require a shared phase reference.

\section{Results}

\textbf{System.} To generate a single photon, we consider a coherently driven circuit quantum electrodynamical (cQED) setup where a superconducting transmon qubit is dispersively coupled to a 3D microwave cavity with a resonance frequency $\omega_\textrm{c}/2\pi =$ \SI{10.619}{\giga\hertz}. The dynamics of the system are described in the rotating frame of the drive by the Hamiltonian
\begin{equation}
\begin{split}
\mathcal{H}/\hbar = &(\omega_{\rm c} - \omega_{\rm d}) a^\dag a + (\omega_{\rm{ge}} - \omega_{\rm d}) b^\dag b + \frac{\alpha}{2} b^\dag b^\dag b b  \\
&+ g (a^\dag b + a b^\dag) + \frac{1}{2}\left[\Omega(t)a^\dag + \Omega^*(t)a\right].
\end{split}
\label{eqn:ham1}
\end{equation}
The qubit is coupled to the cavity with coupling strength $g/2\pi = $ \SI{156.1}{\mega\hertz} and it is driven by coherent microwaves at frequency $\omega_\textrm{d}$ with time-dependent complex amplitude $\Omega(t)$ through the cavity. In Eq.~\eqref{eqn:ham1}, $a$ and $b$ are defined as the cavity and transmon annihilation operators, and $\omega_{\textrm{ge}}/2\pi =$ \SI{7.813}{\giga\hertz} is the qubit $|\textrm{g}\rangle$--$|\textrm{e}\rangle$ transition frequency separated from the $|\textrm{e}\rangle$--$|\textrm{f}\rangle$ transition frequency $\omega_{\textrm{ef}} = \omega_{\textrm{ge}} + \alpha$ by the transmon anharmonicity $\alpha/2\pi =$ \SI{-340}{\mega\hertz}. The cavity and qubit are dispersively coupled, i.e., $\abs{\omega_\textrm{ge} - \omega_\textrm{c}} \gg g$, which allows us to read out the qubit state based on the qubit-state-dependent dispersive shift of the cavity resonance frequency. The cavity is coupled to an external transmission line with an external coupling rate $\kappa_{\textnormal{ex}}/2\pi =$ \SI{2.91}{\mega\hertz}. The relaxation and coherence times between the $|\textrm{g}\rangle$--$|\textrm{e}\rangle$ and $|\textrm{e}\rangle$--$|\textrm{f}\rangle$ states are $T_1^\textrm{ge} =$ \SI{26}{\micro\second}, $T_1^\textrm{ef} =$ \SI{15}{\micro\second} and $T_2^\textrm{ge} =$ \SI{15}{\micro\second}, $T_2^\textrm{ef} =$ \SI{16}{\micro\second}, respectively.

\textbf{Dynamics of time-bin qubit generation.} The state of two time-bin modes can be represented in the photon number basis in two orthogonal temporal modes
\begin{equation}
  \left| \psi_\textrm{TL} \right\rangle = \sum_{n,m=0}^\infty C_{nm}|nm\rangle,
\label{eqn:tempmodestate}
\end{equation}
where $|nm\rangle := |n\rangle_\textrm{E}\otimes |m\rangle_\textrm{L}$ represents the photon number states of the earlier (E) and later (L) modes, respectively, with $\sum\nolimits_{n,m=0}^\infty |C_{nm}|^2 = 1$.

The protocol for quantum state transfer from a superconducting qubit to a time-bin qubit is shown in Fig.~\ref{fig:f1}(a). We prepare the superconducting qubit in a superposition state $\alpha_\textrm{q}|\textrm{g}\rangle + \beta_\textrm{q}|\textrm{e}\rangle$ and transfer the state to $\alpha_\textrm{q}|\textrm{e}\rangle + \beta_\textrm{q}|\textrm{f}\rangle$ with a sequence of $\pi_\textrm{ef}$ and $\pi_\textrm{ge}$ pulses at frequencies $\omega_\textrm{ef}$ and $\omega_\textrm{ge}$, respectively. 

We induce the transition between the $|\textrm{f}0\rangle$ and $|\textrm{g}1\rangle$ states of the combined qubit--cavity system with a drive pulse to generate a shaped single photon inside a transmission line~\citep{bib:tunableqccoupling}. The $|\textrm{f}0\rangle$--$|\textrm{g}1\rangle$ transition frequency is defined as $\omega_\textrm{f0g1} = 2\omega_\textrm{ge} + \alpha - \omega_\textrm{c}$. When the drive frequency matches this transition, the microwave-induced effective coupling between $|\textrm{f}0\rangle$ and $|\textrm{g}1\rangle$ can be derived from the system Hamiltonian in Eq.~\eqref{eqn:ham1}
\begin{equation}
  g_\textrm{eff}(t) = \frac{\sqrt{2}\alpha g^2}{4(\omega_\textrm{c} - \omega_\textrm{ge})^3}\Omega(t).
\end{equation}
Here, the complex amplitude $\Omega(t) = \exp[{i\phi(t)}]|\Omega(t)|$ has a phase degree of freedom $\phi(t)$. By applying this coupling pulse to the sample we can generate a photon inside the cavity. The photon in the cavity will decay to the waveguide at the external coupling rate $\kappa_\textrm{ex}$. Thus, the coefficient $\beta_\textrm{q}$ is transferred to the photon in the E mode of the time-bin qubit. The second coefficient, $\alpha_\textrm{q}$, is transferred to the propagating microwave mode by driving the qubit with a $\pi_\textrm{ef}$ pulse and the coupling pulse once afterwards. 

If the generation protocol has ideal efficiency, the coefficients $\alpha_\textrm{q}$ and $\beta_\textrm{q}$ are transferred to the modes $|01\rangle$ and $|10\rangle$ as $C_\textrm{01} = \alpha_\textrm{q}$ and $C_\textrm{10} = \beta_\textrm{q}$. Since the original qubit state is normalized, $\abs{C_\textrm{01}}^2 + \abs{C_\textrm{10}}^2 = 1$, and all of the other coefficients in Eq.~\eqref{eqn:tempmodestate} become zero. Thus, the transfer process of the qubit state to propagating microwave mode in the temporal mode basis represents the mapping $\alpha_\textrm{q}|\textrm{g}\rangle + \beta_\textrm{q}|\textrm{e}\rangle \mapsto \alpha_\textrm{q}|01\rangle + \beta_\textrm{q}|10\rangle$. We can therefore define the temporal modes $|01\rangle \equiv |\mathrm{L}\rangle$ and $|10\rangle \equiv |\mathrm{E}\rangle$ as the basis states of a dual-rail time-bin qubit. One should note that the time-bin qubit basis states have a single photon, meaning that a valid qubit state can be confirmed with a parity measurement of the total photon number in the two temporal modes.


\textbf{Characterization of the experimental setup.} A schematic of the experimental configuration for generating and measuring the propagating time-bin qubit state is shown in Fig.~\ref{fig:f1}(b). We input the qubit control pulses, qubit state readout pulse, and coupling pulse, to the cavity cooled down to \SI{30}{\milli\kelvin} inside a dilution refrigerator. We amplify the generated time-bin qubit signal with a flux-driven Josephson parametric amplifier (JPA) operated in the degenerate mode by driving the JPA with two successive microwave pulses at frequency $\omega_\textrm{p} = 2\omega_\textrm{c}^\textrm{g}$ where $\omega^\textrm{g}_\textrm{c}/2\pi =$ \SI{10.628}{\giga\hertz} is the dressed cavity frequency when the qubit is in the ground state. The measured signal is demodulated with a local oscillator at frequency $\omega_\textrm{c}^\textrm{LO}$ shifted from $\omega_\textrm{c}^\textrm{g}$ by the sideband frequency $-2\pi \times 50$ \si{\mega\hertz}.

%
\begin{figure}
\begin{center}
  \includegraphics[width=80mm]{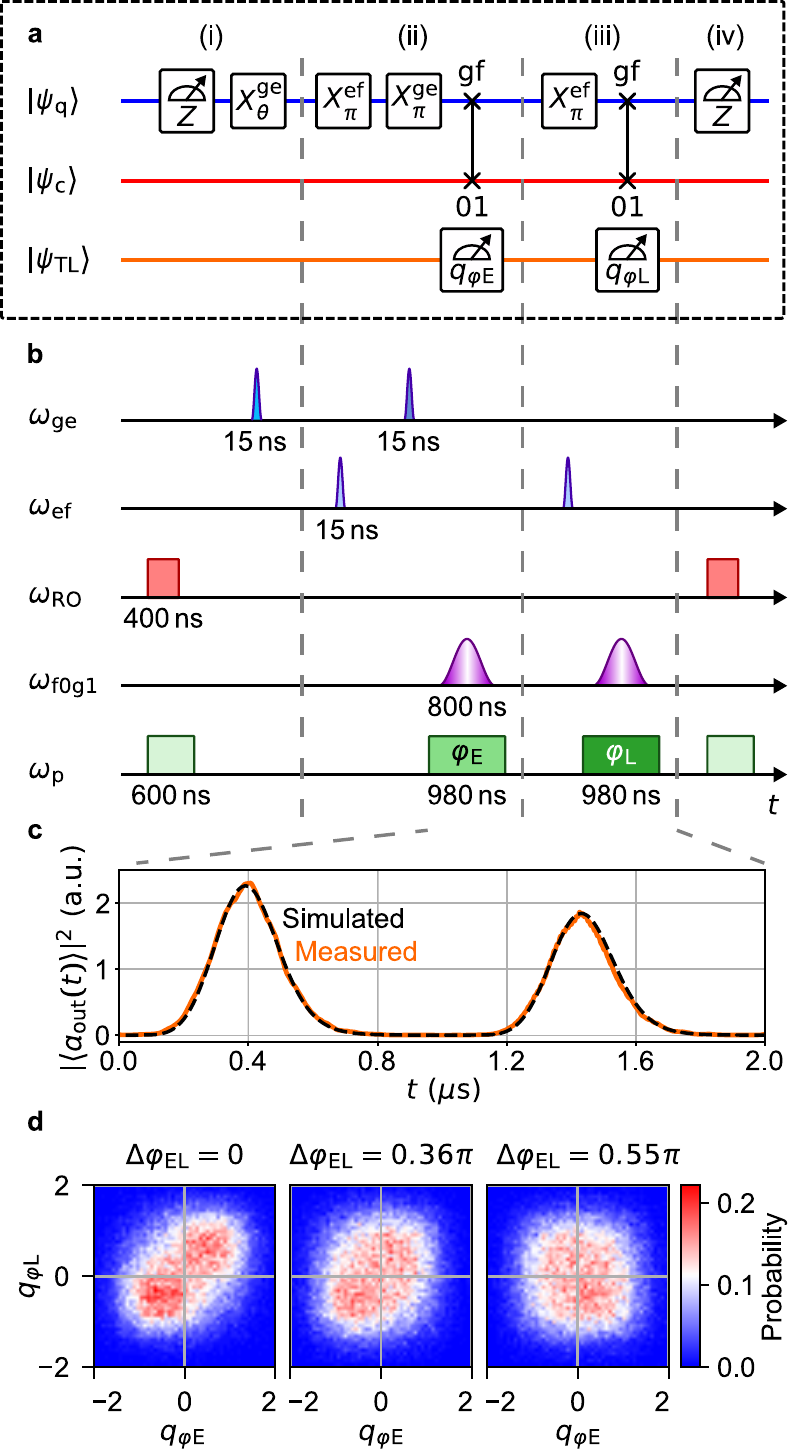} 
\caption{
\textbf{Time-bin qubit generation and characterization sequence.} (a)~Quantum circuit representation of time-bin qubit generation. (i)~Preparation of an arbitrary transmon qubit state. (ii)~Transfer of the transmon qubit state to the first temporal mode of a time-bin qubit and measurement of the first quadrature. (iii)~Transfer of the remaining transmon qubit population to the second temporal mode and measurement of the second quadrature. (iv)~The process ends with a measurement of the qubit state.
(b)~Corresponding time-domain pulse sequence. The pulses are generated at five different frequencies depicted in Fig.~\ref{fig:f1}. The JPA pump pulses have phases $\varphi_\textrm{E}$ and $\varphi_\textrm{L}$. (c)~Measured (orange line) and numerically simulated (black dashed line) mean field amplitude squared of the generated time-bin qubit. (d)~Measured distribution of quadratures $(q_{\varphi\textrm{E}}, q_{\varphi\textrm{L}})$ for the two temporal modes of a time-bin qubit prepared in state $(1/\sqrt{2})(| \textrm{L} \rangle + | \textrm{E} \rangle)$. Each of the three distributions correspond to 72637 post-selected samples measured for quadratures with phase difference $\Delta\varphi_\textrm{EL} = \varphi_\textrm{E} - \varphi_\textrm{L}$ conditioned on the transmon qubit being in the ground state both before and after the time-bin generation.
} 
  \label{fig:f2} 
\end{center}
\end{figure}

We estimate the measurement efficiency for our generation and characterization system by measuring the marginal distribution along a given quadrature in phase space and reconstructing the Wigner function of a single-rail single-photon state $|1\rangle$ in Figs.~\ref{fig:f1}(c)-(d). We only consider measurements where the qubit is in the ground state both before and after the measurement. In the marginal distribution of the measured signal, we extract from a theoretical fit~\citep{bib:fitprobdens} a single photon probability of $P_{|1\rangle} = 0.591 \pm 0.038$ with 95\% confidence intervals. We obtain a fidelity of $0.556 \pm 0.009$ for the reconstructed Wigner function and observe a negative region in the quasiprobability distribution near the origin of the phase space~[Fig.~\ref{fig:f1}(d)], demonstrating negativity of the measured state without loss-correction for detection inefficiency. We define the error interval of the fidelity as three times the standard deviation obtained from bootstrapping~\citep{bib:bootstrap} of the tomography data. We obtain from an analytical calculation (see Supplementary Information~\citep{bib:supplementary}) the possible maximum generation efficiency of $\eta_\textrm{gen}= 0.83 \pm 0.02$ with the parameters in our system, resulting in the minimum measurement efficiency of $\eta_\textrm{meas} = 0.67 \pm 0.01$, comparable to recent experiments in similar systems~\citep{bib:optdispersive, bib:besse2018single, bib:measeffkindel} and mostly explained by the insertion loss of the circulators and isolators.

\textbf{Quadrature distribution of microwave time-bin qubit signal.} The pulse sequence used in the experiment for time-bin qubit generation is shown as a quantum circuit in Fig.~\ref{fig:f2}(a) and as temporal waveforms with different angular frequencies in Fig.~\ref{fig:f2}(b). We perform a $z$-basis dispersive readout on the qubit state~\citep{bib:dispersive,bib:optdispersive} with an assignment fidelity of 0.99 to initialize the qubit, and at the end of the generation sequence to measure whether the transfer sequence results in the qubit being in the ground state or not.

In Fig.~\ref{fig:f2}(c), we show the measured and simulated mean field amplitude squared $|\langle a_\textrm{out}(t)\rangle|^2$ of the state $(1/\sqrt{2})|00\rangle + (1/2)|10\rangle + (1/2)|01\rangle$ as a function of time. The magnitude is calculated according to the theory in the Supplementary Information~\citep{bib:supplementary}. The measured amplitude is normalized to match the simulated amplitude by defining that the integrals calculated over the time interval for the squared amplitudes must be equal. We only consider here measurement events where the transmon qubit was measured as being in the ground state both before and after the generation sequence. We utilize the shape of the measured temporal mode amplitudes to calculate the quadrature distributions of the time-bin qubit. The correlation between the measurements changes based on the selected quadratures, as shown in Fig.~\ref{fig:f2}(d).

%
\begin{figure}[t]
\begin{center}
  \includegraphics[width=80mm]{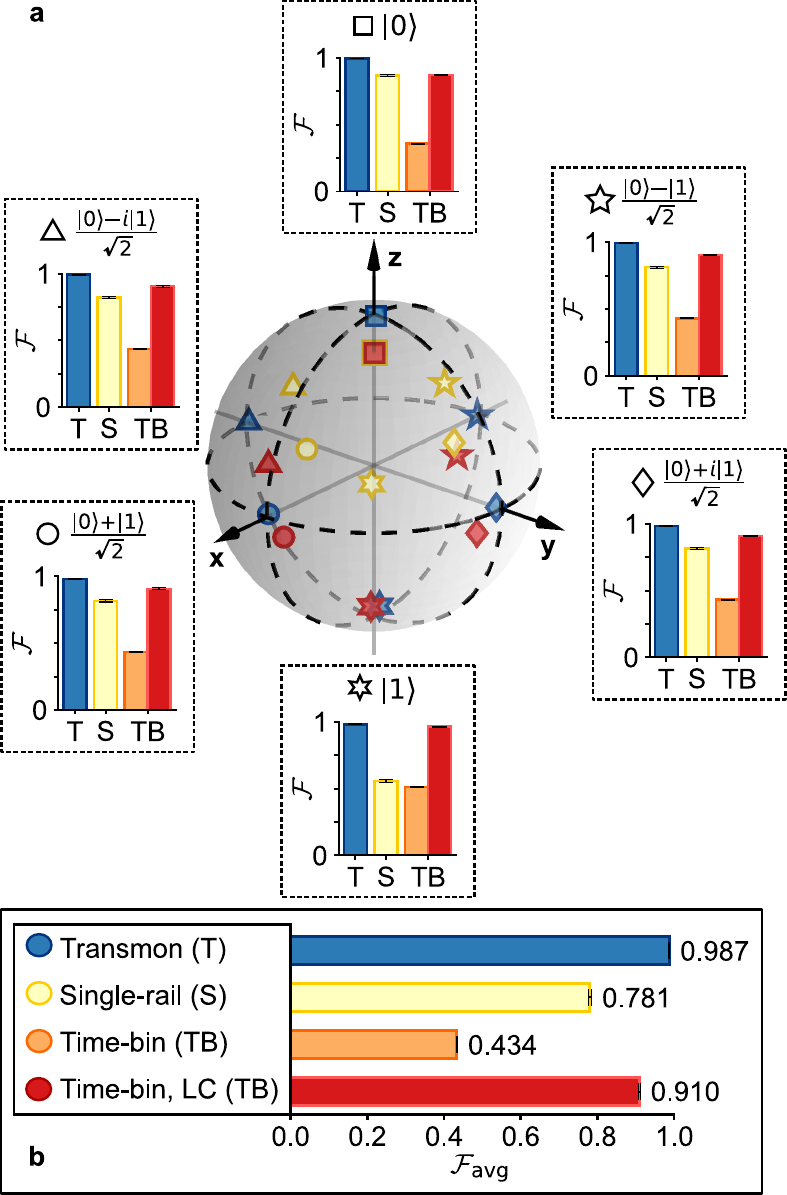} 
\caption{
\textbf{Reconstructed qubit states.} (a)~Reconstructed six cardinal states and their state preparation fidelities measured for a transmon qubit (T)~(basis: $|\textrm{g}\rangle$ and $|\textrm{e}\rangle$), a photonic number-basis qubit (S)~(basis: $|0\rangle_\textrm{S}$ and $|1\rangle_\textrm{S}$), and a time-bin qubit (TB)~(basis: $|\textrm{L}\rangle$ and $|\textrm{E}\rangle$). The states $|0\rangle$ and $|1\rangle$ correspond to the basis states of each qubit in the respective order. The time-bin qubit fidelity is shown both without and with photon-loss correction (LC). The measured transmon qubit, number-basis qubit, and loss-corrected time-bin qubit states are shown inside the Bloch sphere. (b)~Average fidelity calculated over all of the six states for each qubit. The error estimates correspond to three times the standard deviation (99.7\% confidence interval) obtained from bootstrapping of the tomography data.
} 
  \label{fig:f3}
\end{center}
\end{figure}
%
%

\textbf{Characterization of microwave photonic qubit states.} We experimentally prepare the transmon, single-rail number-basis and time-bin qubits in the six cardinal states of the Bloch sphere, as shown in Fig.~\ref{fig:f3}(a). We define the number-basis qubit state basis as $|0\rangle\equiv|0\rangle_\textrm{S}$ and $|1\rangle\equiv|1\rangle_\textrm{S}$, corresponding to no excitation or a single excitation in a single mode. The number-basis qubit states are generated with a sequence similar to the time-bin generation sequence in Fig.~\ref{fig:f2}, but with only the first two qubit control pulses and the first coupling and JPA pump pulses. A series of qubit-state readouts along the three Bloch sphere axes are performed to reconstruct the transmon qubit state. All of the measurements are performed in single-shot. 

We calculate the fidelity of each prepared state as
\begin{equation}
  \mathcal{F} = \langle \psi_\textrm{t} | \rho | \psi_\textrm{t} \rangle,
\end{equation}
where the pure target state is defined as $| \psi_\textrm{t} \rangle$ and $\rho$ is the measured qubit state.

\textbf{Transmon qubit tomography.} For the transmon qubit states, we only consider measurement events where the qubit is initially measured to be in the ground state. On average, 87.5\% of our measurement events fulfill this condition. Given the above condition, we measure a state preparation fidelity of $\mathcal{F}^\textrm{avg}_\textrm{T} = 0.987 \pm 0.001$ averaged over the six cardinal states~[Fig.~\ref{fig:f3}(b)], limited mainly by the qubit control pulse fidelity and readout assignment fidelity.

\textbf{Single-rail number-basis qubit tomography.} For the single-rail states, we post-select the measurement events where both of the readouts before and after the generation sequence result in the qubit state being assigned to the ground state. On average, we keep 82.6\% of all data in the tomography process.

We prepare the single-rail number-basis qubit states with a fidelity of $\mathcal{F}^\textrm{avg}_\textrm{SR} = 0.781 \pm 0.003$, noticeably lower than the transmon qubit states. The difference in fidelity is caused by relaxation and dephasing of the transmon qubit state during single photon generation and photon loss during photon transfer from the qubit to the JPA and heterodyne detector. The effect of photon loss can be observed in the Bloch sphere as a bias towards the $|0\rangle$ state for all of the six cardinal states.

\textbf{Time-bin qubit tomography.} We post-select the time-bin measurement events where both of the readouts result in the transmon qubit being in the ground state corresponding to 80.4\% of all measurements. We discuss the other measurement events in more detail in the Supplementary Information~\citep{bib:supplementary}.

Without loss-correction, we measure an average state-preparation fidelity of $\mathcal{F}^\textrm{avg}_\textrm{TB} = 0.434 \pm 0.001$. Since the generation sequence is longer than that of the single-rail qubit, the effect of qubit control pulse infidelity and qubit dephasing and relaxation on the state preparation fidelity also becomes stronger. Furthermore, we emulate an effective parity measurement on the time-bin qubit density matrices by selecting the density-matrix elements which correspond to the single-photon subspace. We then normalize the time-bin basis density matrices obtained through the selection and obtain a loss-corrected time-bin qubit average state fidelity of $\mathcal{F}^\textrm{avg}_\textrm{TB, LC} = 0.910 \pm 0.002$.
 
%
\begin{figure}[t]
\begin{center}
  \includegraphics[width=80mm]{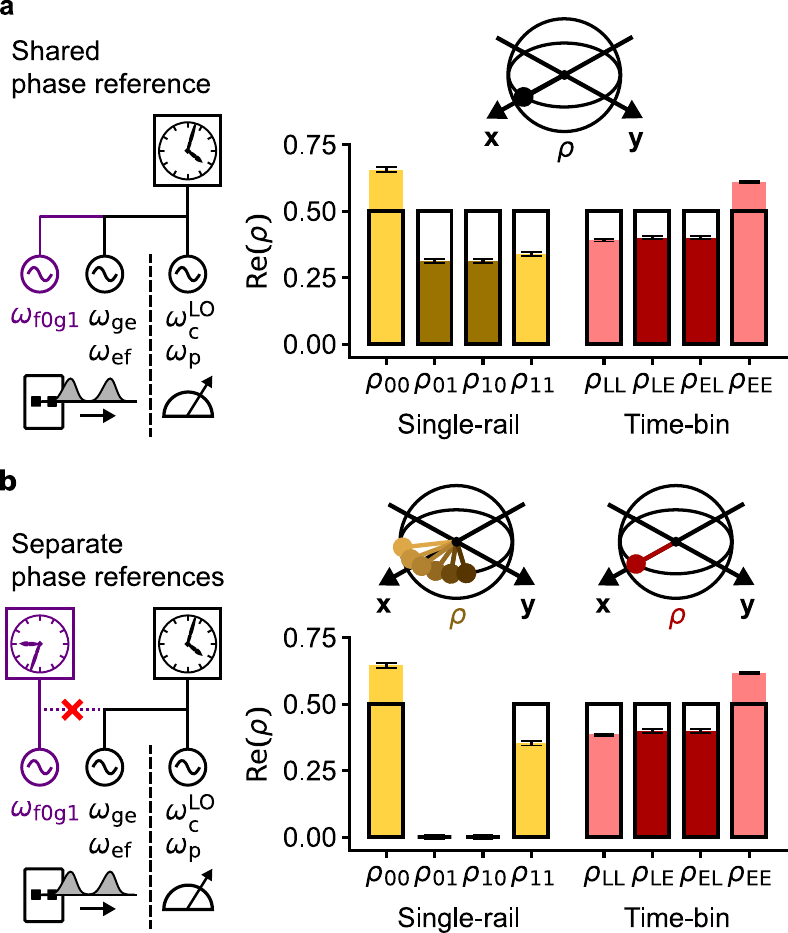} 
\caption{
\textbf{Effect of the lack of a shared phase reference on photonic qubit generation.} Measured real parts of density matrix elements for a single-rail number-basis qubit~(basis: $|0\rangle_\textrm{S}$ and $|1\rangle_\textrm{S}$)  and a loss-corrected time-bin qubit~(basis: $|\textrm{L}\rangle$ and $|\textrm{E}\rangle$). In (a), all of the microwave sources used share the same reference clock (phase reference). In (b), there is no shared reference clock. The Bloch spheres refer to the stability of the phase between each measured state. In the bar plots, the rectangles indicate the ideal values and the error bars are obtained from bootstrapping.
} 
  \label{fig:f4}
\end{center}
\end{figure}

\textbf{Phase robustness of the time-bin qubit.} We measure and reconstruct the density matrices of the single-rail qubit and time-bin qubits for the coherent superposition states $(1/\sqrt{2})(|0\rangle_\textrm{S} + |1\rangle_\textrm{S})$ and $(1/\sqrt{2})(|\textrm{L}\rangle + |\textrm{E}\rangle)$ when the photon source does and does not share the same relative phase reference with the detector, as shown in Figs.~\ref{fig:f4}(a) and (b), respectively. To experimentally realize this condition, we use a separate reference clock for the microwave source which generates the coupling pulse carrier signal than for the other two microwave sources used for qubit control, JPA operation, and demodulation of single-photon signal.

When all of the microwave sources share the same external rubidium clock [Fig.~\ref{fig:f4}(a)], phase coherence is maintained between the generated photons, and the tomography results in a single-rail qubit state fidelity of $\mathcal{F}^\textrm{shared}_\textrm{SR} = 0.811 \pm 0.007$ and time-bin qubit state fidelity of $\mathcal{F}^\textrm{shared}_\textrm{TB} = 0.901 \pm 0.006$. The time-bin qubit is slightly more coherent than the single-rail qubit since the phase reference drifts very slowly even with a shared external clock. In Fig.~\ref{fig:f4}(b) we disconnect the microwave source for the coupling pulse from the shared clock. Due to the phase drift between the two clocks, the single photon signal generated by the coupling pulse has a different phase reference each time. Thus, as we observe in the measured off-diagonal matrix elements, the measured single-rail qubit is dephased completely, resulting in a single-rail preparation fidelity of $\mathcal{F}^\textrm{sep}_\textrm{SR} = 0.500 \pm 0.008$. In contrast, for the time-bin qubit, the phase information is not lost since the relative phase between the two temporal modes determines the phase information of the qubit, resulting in a time-bin qubit state fidelity of $\mathcal{F}^\textrm{sep}_\textrm{TB} = 0.899 \pm 0.006$.

\section{Discussion}


We successfully performed on-demand generation of microwave time-bin qubits by driving a 3D circuit-QED system in dispersive regime and characterized the resulting quantum states with maximum likelihood estimation of two-mode signal amplified by a JPA in heterodyne measurement. Our tomography method allowed us to perform Wigner tomography of a general two temporal mode microwave state with a single detector by switching the measurement quadrature in time between the two temporal modes. We measured an average time-bin qubit state preparation fidelity of 0.910 after loss correction. We also demonstrated that the phase information of the time-bin qubit is stored in the relative phase of the temporal modes and that the lack of a shared phase reference does not cause the time-bin qubit to dephase. By performing a quantum non-demolition measurement of the time-bin qubit with the method in Ref.~\citep{bib:kono} or Ref.~\citep{bib:besse2018single}, it is possible to perform loss-correction on the time-bin qubits in real time to realize robust information transfer and distributed computation in a superconducting qubit network. Our tomography method can also be extended to microwave cluster states with an arbitrary number of temporal modes without any additional detector hardware by adding a new JPA pump pulse for each additional temporal mode. The quadrature detection can be used to realize remote state preparation schemes~\citep{bib:babichev2004remote}. It is also possible to combine our method with an entanglement witness~\citep{bib:morin2013witnessing} or select the measurement quadratures adaptively to characterize the entanglement and state with minimal number of measurements.

\section{Methods}

\textbf{Pulse calibration.} We define the qubit control pulses as Gaussian-shaped pulses while the shape of the coupling pulse is defined as a cosine pulse $[1 - \cos(2\pi t/w)]/2$ with width $w$ for $t \in [0, w]$. We optimize the width, separation, amplitude, phase, and frequency of all pulses in parameter sweep experiments by maximizing the assignment and state preparation fidelities for each parameter separately~\citep{bib:supplementary}. In addition, we apply DRAG~\citep{bib:drag} to the qubit control pulses and chirp the coupling pulses to limit the effect of the $|\textrm{f}\rangle$-state Stark shift affecting the phase of the generated photon wave packet~\citep{bib:shapedphotons}. We also add a constant phase shift to the second coupling pulse relative to the first to reduce the effect of $|\textrm{e}\rangle$-state Stark shift. The optimization of the coefficients for DRAG and chirping is detailed in the Supplementary Information~\citep{bib:supplementary}.

\textbf{Wigner tomography of two temporal modes.} We use JPA phase-sensitive amplification together with heterodyne measurement to reconstruct the quantum state of the single-rail number-basis and time-bin qubits with iterative maximum likelihood estimation performed on measured quadrature distributions of the temporal modes~\citep{bib:tomoreview, bib:maximumlikelihood}. For two temporal modes, we sweep the JPA pump pulse phases for two different angles, $\varphi_\textrm{E} \in [0, 2\pi]$ and $\varphi_\textrm{L} \in [0, 2\pi]$, to change the quadrature of amplification independently for each mode. We measured a JPA gain of \SI{26.8}{\decibel} for the single photon signal (See Supplementary Information~\citep{bib:supplementary}). To reduce the amount of measurements, we perform the tomography for $12 \times 12$ different quadratures in the two modes. For each quadrature pair, we measure $10^4$ samples.

\textbf{Bootstrapping of the reconstructed density matrices.}
We estimate the error of the reconstructed density matrices and state preparation fidelity of the qubit states by bootstrapping the tomography measurement events~\citep{bib:bootstrap}. We resample the data measured for each qubit state and perform maximum likelihood estimation on the resampled data set to obtain a bootstrapped density matrix. By performing this procedure for a number of bootstrapping samples, we can calculate the distribution and standard deviation of the reconstructed density matrix elements. We use 250 bootstrapping samples for each tomography measurement.

\section{Acknowledgements}

We acknowledge fruitful discussions with Y. Tabuchi, M. Fuwa, D. Lachance-Quirion, and N. Gheeraert. This work was supported in part by UTokyo ALPS, JSPS KAKENHI (No.~16K05497 and No.~26220601), JST ERATO (No.~JPMJER1601), and MEXT Quantum Leap Flagship Program (MEXT Q-LEAP No.~JPMXS0118068682). J.I. was supported by the Oskar Huttunen Foundation.

\section{Author contributions}

J.I., S.K., S.Y., and M.K. designed and performed the experiments. J.I., S.K., and Y.S. analysed the data. S.K. and S.Y. fabricated the device. K.K. performed the analytical calculations. J.I. wrote the manuscript with feedback from the other authors. Y.N. supervised the project.

\bibliography{bibnew}

\end{document}


\title{Supplementary Information for ``On-demand generation and characterization of a microwave time-bin qubit"}

\author{J. Ilves$^{1}$}
\author{S. Kono$^{1}$}
\author{Y. Sunada$^{1}$}
\author{S. Yamazaki$^{1}$}
\author{M. Kim$^{1}$}
\author{K. Koshino$^{2}$}
\author{Y. Nakamura$^{1,3}$}%

\affiliation{$^{1}$Research Center for Advanced Science and Technology (RCAST), The University of Tokyo, Meguro-ku, Tokyo 153-8904, Japan}
\affiliation{$^{2}$College of Liberal Arts and Sciences, Tokyo Medical and Dental University, Ichikawa, Chiba 272-0827, Japan}
\affiliation{$^{3}$Center for Emergent Matter Science (CEMS), RIKEN, Wako, Saitama 351-0198, Japan}

\date{December 6, 2019}


\maketitle


\setcounter{equation}{0}
\setcounter{figure}{0}
\setcounter{table}{0}
\makeatletter
\renewcommand{\theequation}{S\arabic{equation}}
\renewcommand{\thefigure}{S\arabic{figure}}
\renewcommand{\bibnumfmt}[1]{[S#1]}
\renewcommand{\citenumfont}[1]{S#1}

\section*{S1. Experimental setup}

\begin{figure}[bp]
\bec
  \includegraphics[width=113mm]{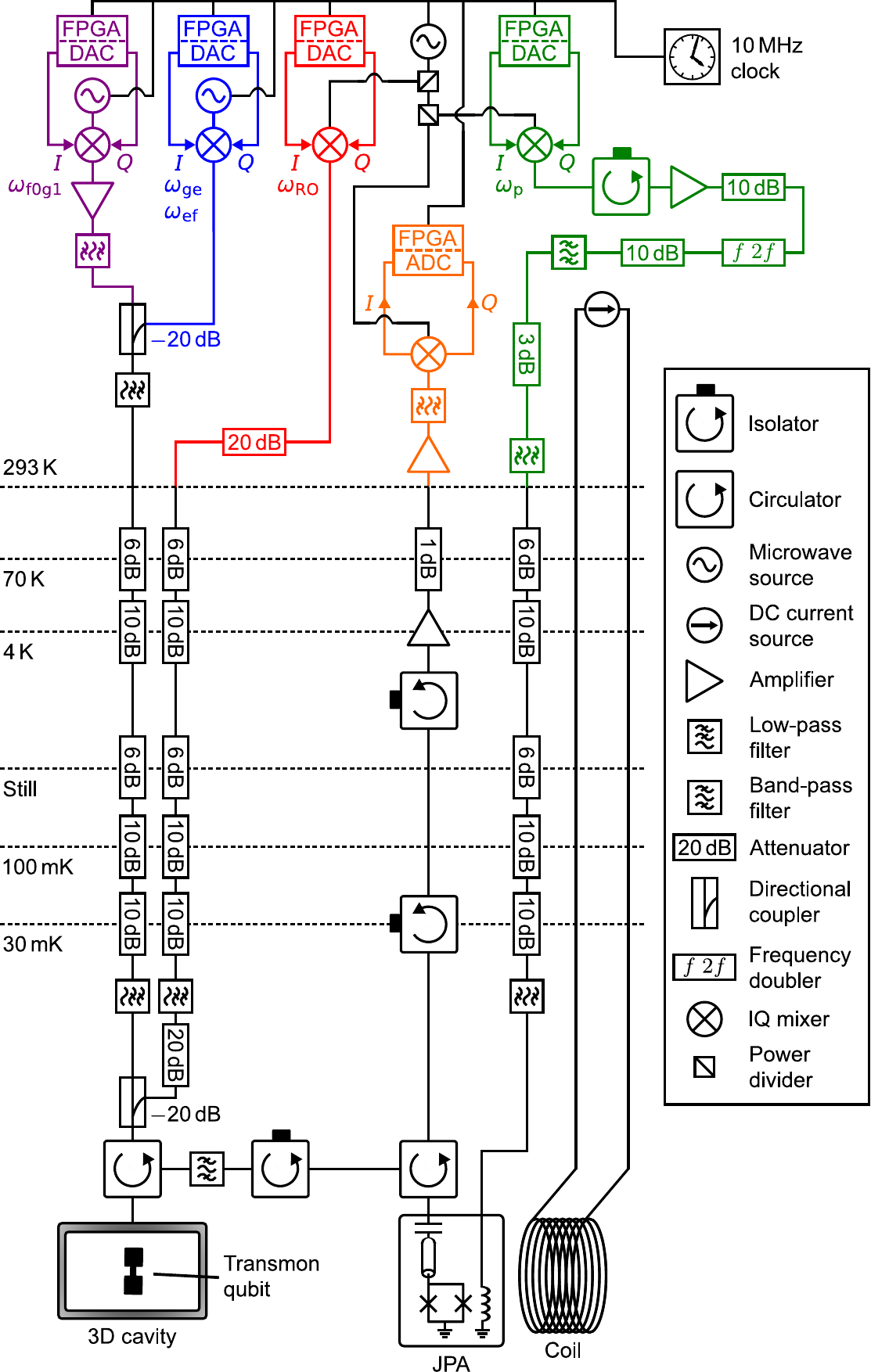} 
\caption{
Schematic of the experimental setup.
}
  \label{fig:setup}
\enc
\end{figure}

A detailed schematic of the experimental devices and configuration used in the measurements inside and outside the dilution refrigerator is shown in Fig.~\ref{fig:setup}. To generate arbitrary waveform pulses at microwave frequencies, we mix low-frequency signal generated at 1-\si{\giga\hertz} sampling rate by the two output channels of field-programmable gate arrays (FPGAs) with high-frequency signal generated by three separate microwave sources acting as local oscillators. The microwave sources and FPGAs are phase-locked by the same 10-\si{\mega\hertz} rubidium clock. We use four different FPGA boards to generate the qubit control pulses, cavity readout pulses, $|\rm{f}0\rangle$--$|\rm{g}1\rangle$ coupling pulse, and pump pulses for a flux-driven Josephson parametric amplifier (JPA) in the experiment. Microwave signal originated at the sample is measured at 1-\si{\giga\hertz} sampling rate after demodulation and analog-to-digital conversion with a fifth FPGA board. In order to keep the phase of the single photon signal coherent over multiple measurements, the local oscillator frequencies of the three microwave sources, $\omega_\textrm{c}^\textrm{LO}$, $\omega_\textrm{ge,ef}^\textrm{LO}$, and $\omega_\textrm{f0g1}^\textrm{LO}$ need to match the condition
\begin{equation}
2\omega_\textrm{ge,ef}^\textrm{LO} - \omega_\textrm{c}^\textrm{LO} - \omega_\textrm{f0g1}^\textrm{LO} = N_r \omega_\textrm{rep},
\end{equation}
where $N_r$ is any integer and $\omega_\textrm{rep}/2\pi$ is the repetition frequency of the measurements.

We use a directional coupler to input the readout pulse through a separate line with additional attenuation to reduce noise at cavity frequency. The transmon qubit consisting of an $\textrm{Al}/\textrm{Al}_2\textrm{O}_3/\textrm{Al}$ Josephson junction fabricated on a sapphire substrate is placed inside a three-dimensional~(3D) aluminum cavity. We tune the frequency of the JPA to match the cavity frequency by appling a DC magnetic field into its SQUID loop. System parameters measured for the sample in frequency- and time-domain measurements are shown in Table~\ref{tbl:params}.

\begin{table}[tp]
\bec
\caption{Measured system parameters.}
  \begin{tabular}{lcl} \hline \hline
\textbf{3D cavity} \\
\hline
Bare frequency & $\omega_{\textnormal c}/2\pi$ & \SI{10.619}{\giga\hertz} \\
Dressed frequency for $\ket{\textnormal{g}}$ & $\omega_{\textnormal c}^{\textnormal g}/2\pi$ & \SI{10.628}{\giga\hertz} \\ 
External coupling rate & $\kappa_{\textnormal{ex}}/2\pi$ & \SI{2.91}{\mega\hertz} \\ 
Internal coupling rate & $\kappa_{\textnormal{in}}/2\pi$ & \SI{346}{\kilo\hertz} \\ \\
\textbf{Transmon qubit} \\
\hline
$\ket{\textnormal{g}}$--$\ket{\textnormal{e}}$ transition frequency & $\omega_{\textnormal{ge}}/2\pi$ & \SI{7.813}{\giga\hertz} \\ 
$\ket{\textnormal{e}}$--$\ket{\textnormal{f}}$ transition frequency & $\omega_{\textnormal{ef}}/2\pi$ & \SI{7.473}{\giga\hertz} \\
Anharmonicity & $\alpha/2\pi$ & \SI{-340}{\mega\hertz} \\
$\ket{\textnormal{g}}$--$\ket{\textnormal{e}}$ energy relaxation time & $T_1^{\textnormal{ge}}$ & \SI{26}{\micro\second} \\ 
$\ket{\textnormal{g}}$--$\ket{\textnormal{e}}$ coherence time & $T_2^{\textnormal{ge}}$ & \SI{15}{\micro\second} \\
$\ket{\textnormal{e}}$--$\ket{\textnormal{f}}$ energy relaxation time & $T_1^{\textnormal{ef}}$ & \SI{15}{\micro\second} \\ 
$\ket{\textnormal{e}}$--$\ket{\textnormal{f}}$ coherence time & $T_2^{\textnormal{ef}}$ & \SI{16}{\micro\second} \\ \\
Qubit--cavity coupling rate & $g/2\pi$ & \SI{156.1}{\mega\hertz}\\
Maximum $|\rm{f}0\rangle$--$|\rm{g}1\rangle$ coupling rate & $g_\textrm{eff}^\textrm{max}/2\pi$ & \SI{2.2}{\mega\hertz}\\ \\
\hline \hline
  \end{tabular}\
\label{tbl:params}
\enc  
\end{table}

\section*{S2. Theoretical description of time-bin qubit wave packet generation}

\begin{figure}[t]
\begin{center}
  \includegraphics[width=141.386mm]{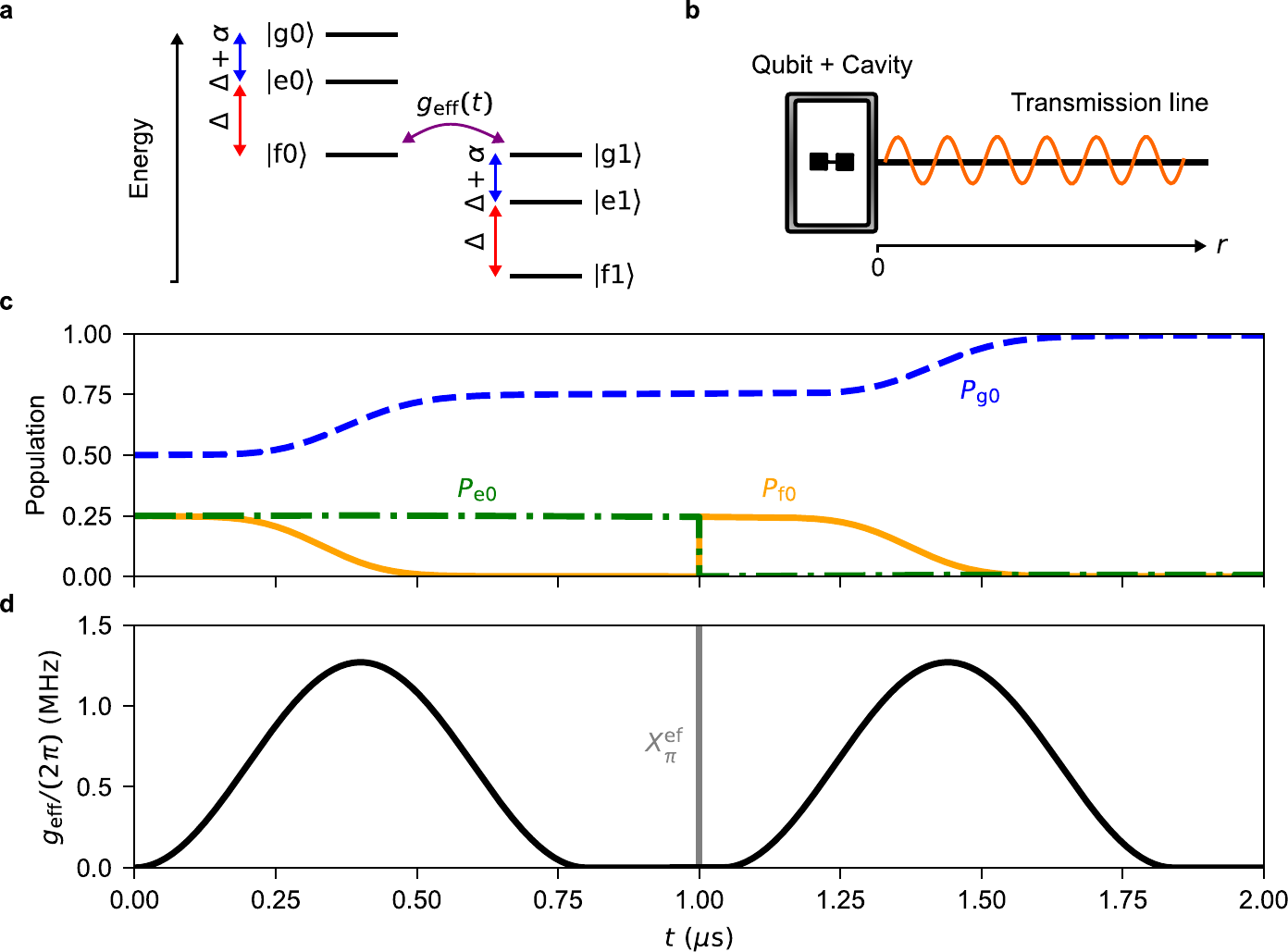} 
\caption{
\textbf{Analytical model for time-bin qubit generation.} (a)~Energy level diagram of the qubit--cavity system in a frame rotating at the drive frequency. (b)~Schematic of the qubit--cavity system coupled to a transmission line.
(c)~Simulated change in qubit--cavity state populations as a function of time during time-bin qubit generation.
(d)~Effective coupling between the $\ket{\textrm{f0}}$ and $\ket{\textrm{g1}}$ states defined in the simulation using the parameters in the experiments.
} 
  \label{fig:s2}
\end{center}
\end{figure}

By examining the single-photon generating system dynamics in detail we calculate the shape of the time-bin qubit wave packet generated in a transmission line. We consider a transmon qubit as a three-level system with transition frequency $\omega_\textrm{ge}$ between the $\ket{\textrm{g}}$ and $\ket{\textrm{e}}$ states and anharmonicity $\alpha$. The transmon qubit is coupled to a cavity, which in turn is coupled to the transmission line. The total system Hamiltonian can be written in a frame rotating at frequency $\omega_\textrm{f0g1} = 2\omega_\textrm{ge} + \alpha - \omega_\textrm{c}$ as
\begin{equation}
  \mathcal{H}_\textrm{tot}(t) = \mathcal{H}_\textrm{qc}(t) + \mathcal{H}_\textrm{TL} + \mathcal{H}_\textrm{rel} + \mathcal{H}_\textrm{dep},
  \label{eqn:fullham}
\end{equation}
where $\mathcal{H}_\textrm{qc}(t)$ describes the qubit and cavity and their interaction, $\mathcal{H}_\textrm{TL}$ describes the transmission line, and $\mathcal{H}_\textrm{rel}$ and $\mathcal{H}_\textrm{dep}$ describe the relaxation and dephasing, respectively. Defining $\hbar = v_\textrm{m} = 1$, where $v_\textrm{m}$ is the microwave velocity in the transmission line, the qubit-cavity dynamics due to an effective coupling $g_\textrm{eff}(t)$ between the $\ket{\textnormal{f}0}$ and $\ket{\textnormal{g}1}$ states reduces in the rotating frame to
\begin{equation}
\begin{split}
  \mathcal{H}_\textnormal{qc}(t) &= g_\textnormal{eff}(t)\sigma_\textnormal{f0,g1} + g_\textnormal{eff}^*(t)\sigma_\textnormal{g1,f0} \\ 
  &- (\Delta + \alpha)\sigma_\textnormal{e0,e0} - (2\Delta + \alpha)(\sigma_\textnormal{f0,f0} + \sigma_\textnormal{g1,g1}) \\
  &- (3\Delta + 2\alpha)\sigma_\textnormal{e1,e1} - (4\Delta + 2\alpha)\sigma_\textnormal{f1,f1},
\end{split}
\end{equation}
where the operators $\sigma_{vj,wl} = | vj \rangle\langle wl |$ for $\left\{ v, w \right\} = \left\{\textrm{g}, \textrm{e} \right\}$, 
$\left\{ j,l \right\} = \left\{ 0,1,2, \dots \right\}$, and $\Delta = \omega_\textnormal{ge} - \omega_\textnormal{c}$ is the qubit-cavity detuning. In Fig.~\ref{fig:s2}(a), we show the energy level diagram of the composite system in the rotating frame. The Hamiltonian of the transmission line can be written as
\begin{equation}
  \mathcal{H}_\textrm{TL} = \int\! \textrm{d}k \,\left[kb_k^{\dagger}b_k + \sqrt{\kappa/2\pi}\, \left(a^{\dagger}b_k + b_k^{\dagger}a\right)\right],
\end{equation}
where $b_k$ is the annihilation operator of the propagating mode in the transmission line with wave number $k$, $a$ is the cavity photon annihilation operator, and $\kappa$ is the total cavity coupling strength. We model the relaxation channel as
\begin{equation}
  \mathcal{H}_\textrm{rel} = \int\! \textrm{d}k \,\left[kc_k^{\dagger}c_k + \sqrt{1/2\pi}\, \left(D^{\dagger}c_k + c_k^{\dagger}D\right)\right]
\end{equation}
where we define the dissipation operator $D$ as
\begin{equation}
  D = \sqrt{\gamma_\textrm{ge}}s_\textrm{ge} + \sqrt{\gamma_\textrm{ef}}s_\textrm{ef},
\end{equation}
with relaxation times $T_1^\textrm{ge} = 1/\gamma_\textrm{ge}$ and $T_1^\textrm{ef} =  1/\gamma_\textrm{ef}$ from $\ket{\textrm{e}}$ to $\ket{\textrm{g}}$, and from $\ket{\textrm{f}}$ to $\ket{\textrm{e}}$, respectively. In addition, $c_k$ is the annihilation operator for a bosonic mode with energy $\hbar v_\textrm{m} k$ in the relaxation channel, and $s_{vw}$ is the transition operator of the transmon, $s_{vw} = | v \rangle\langle w |$.

The dephasing of the qubit is described by two additional channels
\begin{equation}
  \begin{split}
  \mathcal{H}_\textrm{dep} &= \int\! \textrm{d}k \,\left[kd_{1,k}^{\dagger}d_{1,k} + \sqrt{\gamma_\textrm{p,1}/\pi}\,s_\textrm{ee}\left(d_{1,k}^{\dagger} + d_{1,k}\right)\right] \\
  &+ \int\! \textrm{d}k \,\left[kd_{2,k}^{\dagger}d_{2,k} + \sqrt{\gamma_\textrm{p,2}/\pi}\,s_\textrm{ff}\left(d_{2,k}^{\dagger} + d_{2,k}\right)\right],
  \end{split}
\end{equation}
in which we define $d_{j,k}$ as the annihilation operators of the relaxation channels for three different pure dephasing coefficients $\gamma_{\textrm{p,}j}$.

By defining the Fourier transform of $b_{k}$ as $b_{r} = (2\pi)^{-1/2}\int\textrm{d}k\exp(ikr)b_k$, we can write the state of the system at a given time as
\begin{equation}
  | \psi(t) \rangle = \sum_{u_\textrm{q}\in\{\textrm{g,e,f}\}} \sum_{u_\textrm{c}\in\{0,1\}}  C_\textrm{qc}(u_\textrm{q}, u_\textrm{c}, t)| u_\textrm{q} u_\textrm{c}, 0 \rangle + \int\! \textrm{d}r \,f(r, t)b^{\dagger}_{r} | \textrm{g0}, 0 \rangle + \cdots,
\label{eqn:state}
\end{equation}
where the dots represent the terms with excitations in the relaxation channel, $C_{\textrm{qc}}$ are defined as complex coefficients related to the probability $|C_{\textrm{qc}}|^2$ for the qubit--cavity system to be in a given state, the third state $\ket{0}$ in the tensor product is the vacuum state of the transmission line, and $f(r, t)$ is the time-bin wave packet complex amplitude at position $r$ from the qubit at time $t$, as shown in the schematic in Fig.~\ref{fig:s2}(b). Given the input-output relation
%
%
\begin{equation} 
  b_r(t) = b_{r-t}(0) - i\sqrt{\kappa_\textnormal{ex}}\, \theta(r)\theta(t-r)a(t - r)
\end{equation}
and initial state of the system
\begin{equation}
  | \psi(0) \rangle = C_0\ket{\textrm{g0}, 0} + C_1\ket{\textrm{e0}, 0} + C_2\ket{\textrm{f0}, 0},
\end{equation}
we can calculate the wave packet shape as $f(r, t) = \langle \psi(t) | b_r | \psi(t) \rangle$.

We introduce a unitary time-evolution operator
\begin{equation}
U(t)=\mathcal{T}\exp\left[-i\int_0^t \textrm{d}t' H_\textrm{tot}(t')\right],
\end{equation}
where $\mathcal{T}$ is the time-ordering operator. We define the notation $\langle S(t) \rangle_0 = \langle\psi(0)| S(t) |\psi(0)\rangle$ for any operator $S$. By writing $f(r,t) = \langle U^{\dagger}(t) b_r(0) U(t) \rangle_0$, we have that
\begin{equation}
\begin{split}
  f(r,t) &= \langle b_r(t) \rangle_0 \\
  &= \begin{cases}
-i\sqrt{\kappa_\textnormal{ex}}\,\langle a(t - r) \rangle_0, & (0<r<t)
\\
0, & (\textrm{otherwise})
\end{cases}
\end{split}
\label{eqn:cases}
\end{equation}
since $b_r(0) |\psi(0) \rangle = 0$. Hereafter, we evaluate the amplitude of the generated time-bin qubit at $r=+0$, corresponding to the qubit's position. From Eq.~\eqref{eqn:cases}, we obtain
\begin{equation}
  f(0, t) = -i\sqrt{\kappa_\textnormal{ex}}\,\langle a(t) \rangle_0 = -i\sqrt{\kappa_\textnormal{ex}}\,\langle \sigma_{\textrm{g0},\textrm{g1}}(t)\rangle_0,
\label{eqn:f0_1}
\end{equation}
where we have omitted the terms $\langle\sigma_{\textrm{e0},\textrm{e1}}(t)\rangle_0$ and $\langle\sigma_{\textrm{f0},\textrm{f1}}(t)\rangle_0$ since they are zero.

The time evolution for any system operator $S$ can be solved from the Heisenberg equation
\begin{equation}
\begin{split}
  \frac{\textrm{d}S}{\textrm{d}t} = i[\mathcal{H}_\textnormal{qc}(t), S] &+ \frac{\kappa}{2}(2a^{\dagger}Sa - Sa^{\dagger}a - a^{\dagger}aS) + \frac{1}{2}(2D^{\dagger}SD - SD^{\dagger}D - D^{\dagger}DS)  \\
  &- \gamma_{\textrm{p},1}[[S, s_\textrm{ee}], s_\textrm{ee}] - \gamma_{\textrm{p},2}[[S, s_\textrm{ff}], s_\textrm{ff}] + \textrm{H.c.}
\end{split}
\label{eqn:heisenberg}
\end{equation}
From here on we denote $C_{vj,wl} = \langle \sigma_{vj,wl}(t) \rangle_0$. After substitution of relevant system operators to \eqref{eqn:heisenberg}, we notice that many of the expectation values are zero. The time-evolution of $C_{\textrm{g0},\textrm{g1}}$ can be solved from the following system of equations
%
\begin{align}
  \frac{\textrm{d}}{\textrm{d}t}C_{\textrm{g0},\textrm{f0}} &= \left[i(2\Delta + \alpha) - 1/T_2^\textrm{ef}\right]C_{\textrm{g0},\textrm{f0}} - ig_\textnormal{eff}(t)C_{g0,g1} \label{eqn:g0f0}\\
  \frac{\textrm{d}}{\textrm{d}t}C_{\textrm{g0},\textrm{g1}} &= [i(2\Delta + \alpha) - \kappa/2] C_{\textrm{g0},\textrm{g1}} - ig^*_\textnormal{eff}(t)C_{\textrm{g0},\textrm{f0}} \label{eqn:g0g1}\\
  \frac{\textrm{d}}{\textrm{d}t}C_{\textrm{e0},\textrm{f0}} &= \left[i\Delta - \left(1/T_2^\textrm{ge} + 1/T_2^\textrm{ef}\right)\right]C_{\textrm{e0},\textrm{f0}} - ig_\textnormal{eff}(t)C_{e0,g1} \label{eqn:e0f0}\\
  \frac{\textrm{d}}{\textrm{d}t}C_{\textrm{e0},\textrm{g1}} &= \left[i\Delta - \left(\kappa/2 + 1/T_2^\textrm{ge}\right)\right]C_{\textrm{e0},\textrm{g1}} - ig^*_\textnormal{eff}(t)C_{e0,f0} \label{eqn:e0g1}\\
  \frac{\textrm{d}}{\textrm{d}t}C_{\textrm{g0},\textrm{e0}} &= \left[i(\Delta + \alpha) - 1/T_2^\textrm{ge}\right]C_{\textrm{g0},\textrm{e0}} + \sqrt{\left(T_1^\textrm{ge}T_1^\textrm{ef}\right)^{-1}}\,C_{\textrm{e0},\textrm{f0}}\label{eqn:g0e0}\\
  \frac{\textrm{d}}{\textrm{d}t}C_{\textrm{f0},\textrm{f0}} &= ig^*_\textnormal{eff}(t)C_{\textrm{g1},\textrm{f0}} - ig_\textnormal{eff}(t)C_{\textrm{f0},\textrm{g1}} - \left(1/T_1^\textrm{ef}\right)C_{\textrm{f0},\textrm{f0}}\label{eqn:f0f0}\\
  \frac{\textrm{d}}{\textrm{d}t}C_{\textrm{g1},\textrm{g1}} &= ig_\textnormal{eff}(t)C_{\textrm{f0},\textrm{g1}} - ig^*_\textnormal{eff}(t)C_{\textrm{g1},\textrm{f0}} - \kappa C_{\textrm{g1},\textrm{g1}}\label{eqn:g1g1}\\
  \frac{\textrm{d}}{\textrm{d}t}C_{\textrm{f0},\textrm{g1}} &= ig^*_\textnormal{eff}(t)(C_{\textrm{g1},\textrm{g1}} - C_{\textrm{f0},\textrm{f0}}) - \left(\kappa/2 + 1/T_2^\textrm{ef}\right) C_{\textrm{f0},\textrm{g1}}\label{eqn:f0g1}\\
  \frac{\textrm{d}}{\textrm{d}t}C_{\textrm{g0},\textrm{g0}} &= \kappa C_{\textrm{g1},\textrm{g1}} + \left(1/T_1^\textrm{ge}\right) C_{\textrm{e0},\textrm{e0}}\label{eqn:g0g0}\\
  \frac{\textrm{d}}{\textrm{d}t}C_{\textrm{e0},\textrm{e0}} &= -\left(1/T_1^\textrm{ge}\right) C_{\textrm{e0},\textrm{e0}} + \left(1/T_1^\textrm{ef}\right) C_{\textrm{f0},\textrm{f0}}\label{eqn:e0e0},
\end{align}
where $T_2^\textrm{ge} = (\gamma_\textrm{ge}/2 + \gamma_{\textrm{p},1})^{-1}$ ($T_2^\textrm{ef} = (\gamma_\textrm{ef}/2 + \gamma_{\textrm{p},2})^{-1}$) is the dephasing time of the qubit between $\ket{\textrm{g}}$ and $\ket{\textrm{e}}$ ($\ket{\textrm{e}}$ and $\ket{\textrm{f}}$).

To simulate the time-bin qubit generation process, we solve the equations numerically for two separate time periods. In the first period, we simulate the generation of the first time-bin. After the transfer, we simulate the $X_\pi^\textrm{ef}$ pulse by swapping the $\ket{\textrm{e0}}$ terms with $\ket{\textrm{f0}}$ in the equations and performing a second calculation simulating the generation of the second time-bin with the initial condition matching the state of the system after the $X_\pi^\textrm{ef}$ pulse.

To simulate generation of coherent time-bin qubit state signal amplitude, we initialize the system state as $C_0 = 1/\sqrt{2}\,, C_1 = C_2 = 1/2$. We show the $g_\textnormal{eff}(t)$ defined in the simulations together with the change in the qubit population as a function of time in Figs.~\ref{fig:s2}(c) and (d) calculated with the parameters for the sample given in Table~\ref{tbl:params}. 

In order to calculate the generation efficiency of the system in the simulation, we prepare the system in the initial state $C_0 = C_2 = 1/\sqrt{2}, C_1 = 0$. We perform the photon generation process for the first temporal mode only. We evaluate the photon generation efficiency as
\begin{equation}
\eta_\textrm{gen} = \frac{1}{1 - P_\textrm{e0}^\textrm{sc}}\int_{0}^{\infty}\!\textrm{d}t\,\frac{|f(0,t)|^2}{|C_0|^2|C_2|^2},
\end{equation}
where we have defined $P_\textrm{e0}^\textrm{sc}$ as the leftover population in the $|\textrm{e0}\rangle$ state at \SI{0.95}{\micro\second}. The leftover population is used to match the conditions in the simulation with the experiments. The scaling imposes the condition that the qubit must be in the ground state both before and after the generation protocol. We do not take the leftover population in the $|\textrm{f0}\rangle$ state into account in the scaling since it is smaller than $P_\textrm{e0}^\textrm{sc}$ by over one order. For a calculated value of $P_\textrm{e0}^\textrm{sc} = 0.02$, we thus obtain a generation efficiency of $\eta_\textrm{gen} = 0.83 \pm 0.02$.

\section*{S3. Optimization of qubit control and readout pulse parameters}



To optimize the qubit readout and control parameters, we first perform a rough optimization of the parameters by using a readout with enough visibility to obtain reasonable resolution. We define the $X^\textrm{ge}_{\pi}$ and $X^\textrm{ef}_{\pi}$ qubit control pulses as Gaussian pulses with a fixed width of \SI{15}{\nano\second}. We define the amplitude and frequency of the pulses with Rabi oscillation and Ramsey measurements, respectively. We optimize the $|\textnormal{f}0\rangle$--$|\textnormal{g}1\rangle$ coupling pulse parameters by measuring the Rabi oscillation between the states as a function of the drive frequency and pulse amplitude, finding the optimal point that maximizes state transfer and effective coupling strength.


For further optimization of the qubit readout pulse, we perform two consecutive readouts of the qubit state and maximize the measured assignment fidelity $(1/2)\left[P(\textnormal{g}|\textnormal{g}) + P(\textnormal{e}|\textnormal{e})\right]$ by sweeping over different readout pulse parameter values. The readout frequency is set constant at the frequency of the dressed cavity when the qubit is in the ground state. We vary the rectangular shape readout pulse width, amplitude, and separation between the two pulses in the assignment fidelity measurement sequence. In addition, since the measurements are performed in single shot, we also sweep over different JPA pump pulse amplitudes, widths, and phases to maximize the assignment fidelity. Since we operate the JPA in the degenerate mode, we use the JPA to perform a measurement of the qubit being either in state `s' or `not s' instead of being able to distinguish between the different qubit states at the same time. Here the state `s' refers to either $\ket{\textrm{g}}$, $\ket{\textrm{e}}$ or $\ket{\textrm{f}}$. Depending on whether we want to measure the $\ket{\textrm{g}}$, $\ket{\textrm{e}}$ or $\ket{\textrm{f}}$ state, we change the direction of the phase to match the correct state. Since the readout is a result of IQ demodulated signal, we also optimize the shape of the demodulation integration weight function to match the readout signal. With these optimizations, we measure an assignment fidelity of 0.99 for our readout with a pulse length of \SI{400}{\nano\second}.


To increase the fidelity of the gate operations for short pulses with relatively high power, we apply the Derivative Removal by Adiabatic Gate (DRAG)~\citep{bib:drag} technique to the qubit control pulses. The DRAG technique reduces leakage to unwanted transitions due to the short high power drive pulses. We can write the amplitude of the generated control pulse shape with drive frequency $\omega_\textrm{d}$ as
\begin{equation}
  A(t) = X(t)\cos{(\omega_\textrm{d} t + \phi_\textrm{g})} + \beta \frac{\textrm{d}X(t)}{\textrm{d}t}\sin{(\omega_\textrm{d} t + \phi_\textrm{g})},
\end{equation}
where $\phi_\textrm{g}$ is the drive phase that controls around what axis in the equatorial plane the qubit state rotates in the Bloch sphere and $\beta$ is the DRAG coefficient. The amplitude without DRAG correction is applied as $X(t)$.

\begin{figure}[t]
\begin{center}
  \includegraphics[width=120mm]{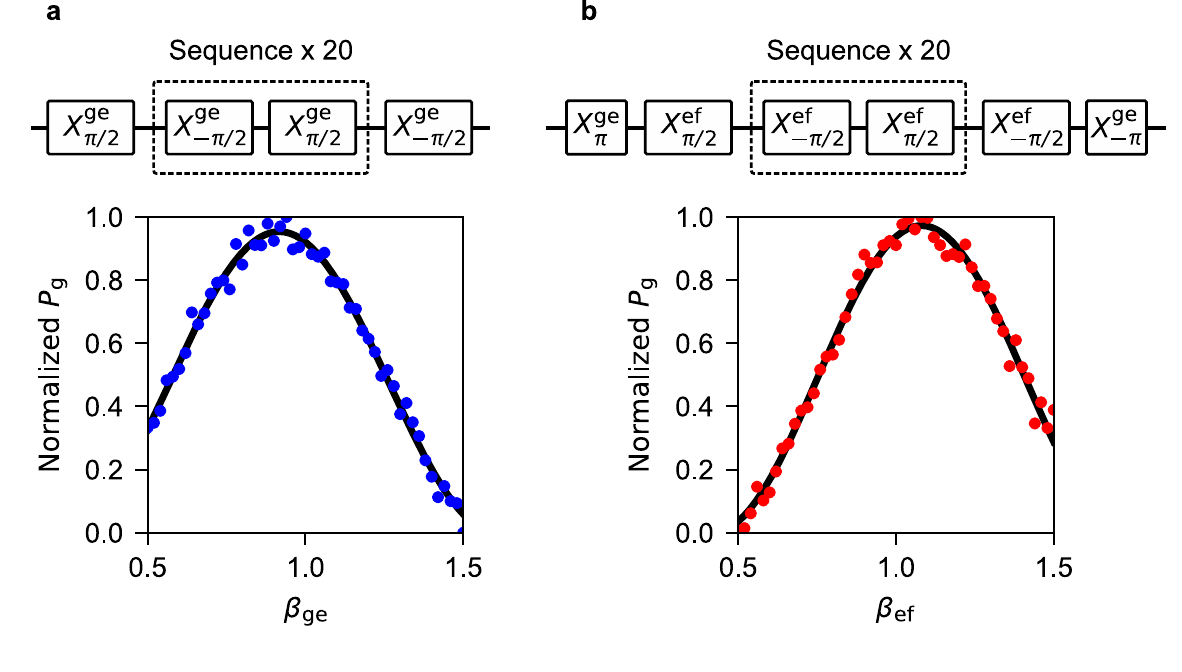} 
\caption{
\textbf{Calibration of DRAG coefficient for qubit control pulses.} (a)~Measured normalized $\ket{\textrm{g}}$ state population during DRAG optimization of a $\pi$ pulse for the $\ket{\textrm{g}}$--$\ket{\textrm{e}}$ transition as a function of the DRAG coefficient $\beta_\textrm{ge}$. The black curve is a cosine function fit to the measured data.
(b)~Same as (a) but for a sequence corresponding to optimizing the DRAG coefficient $\beta_\textrm{ef}$ of a $\pi$ pulse for the $\ket{\textrm{e}}$--$\ket{\textrm{f}}$ transition.
} 
  \label{fig:s3}
\end{center}
\end{figure}

We optimize the DRAG coefficient separately for the $\ket{\textrm{g}}$-$\ket{\textrm{e}}$ and $\ket{\textrm{e}}$-$\ket{\textrm{f}}$ pulses experimentally by measuring the population in the $\ket{\textrm{g}}$ state after a specific control pulse sequence. The sequence for optimizing the $X^\textrm{ge}_\pi$ pulse with DRAG can be written as $X^\textrm{ge}_{\pi/2} (X^\textrm{ge}_{-\pi/2} X^\textrm{ge}_{\pi/2})^N X^\textrm{ge}_{-\pi/2}$, as shown in Fig~\ref{fig:s3}(a), where $N$ corresponds to an integer number describing the number of repetitions of the sequence. In the figure we show measured population in the $\ket{\textrm{g}}$ state normalized to unity as a function of the control pulse amplitude and DRAG coefficient $\beta_\textrm{ge}$ for $N=20$. The sequence amplifies the phase error caused by virtual excitation through undesired states. By tuning the DRAG coefficient $\beta_\textrm{ge}$ to $\beta_\textrm{ge}^\textrm{opt} = 0.92$, we find the optimal point where the population is maximized, corresponding to minimal phase error. A similar sequence can be defined for the $X^\textrm{ef}_\pi$ pulse by exchanging the $X^\textrm{ge}_{-\pi/2}$ and $X^\textrm{ge}_{\pi/2}$ pulses with their $X^\textrm{ef}$ equivalents and by preceding the sequence with a $X^\textrm{ge}_\pi$ pulse reversed after the sequence by a $X^\textrm{ge}_{-\pi}$ pulse. We show the results of this optimization in a similar plot to Fig.~\ref{fig:s3}(a) in Fig.~\ref{fig:s3}(b) with $N=20$ and an optimal point at $\beta_\textrm{ef}^\textrm{opt} = 1.08$ maximizing the population in the ground state. With the optimized parameters for the readout and qubit control pulses, we obtain a probability of 0.979 for the qubit to be measured in the first excited state after initialization by post-selection and performing the $X^\textrm{ge}_{\pi}$ pulse.


\section*{S4. Cancellation of the effect of ac Stark shift on the $|\textrm{f}\rangle$ state}


Due to the ac Stark shift caused by the qubit--cavity excitation swapping pulse, the shape of the generated photon will become distorted. To compensate the effect of the $\ket{\textrm{f}}$-state shift, we generate a chirped pulse $a_\textnormal{c}(t) = a_\textnormal{p}(t)\exp[{i\phi_\textrm{f0g1}(t)}]$ by time-modulating the phase of the original control pulse $a_\textnormal{p}(t)$ with a method based on Ref.~\citep{bib:shapedphotons}. Here, both pulse amplitudes are defined in an arbitrary unit. We apply time-modulation to the phase $\phi_\textrm{f0g1}(t)$ of the drive pulse according to
\begin{equation}
\frac{\textrm{d}\phi_\textrm{f0g1}(t)}{\textrm{d}t} = -\Delta_\textnormal{f0g1}(t),
\end{equation}
where $\Delta_\textnormal{f0g1}(t)$ is the Stark shift of the transition caused by the drive pulse at a given time. Since the ac Stark shift has a quadratic dependence on the pulse amplitude, the general form of the chirped pulse satisfying the above equation reduces to
%
\begin{equation}
  a_\textnormal{c}(t) =  a_\textnormal{p}(t) \exp{\left( -i \int_0^t \textrm{d}t' C_\textrm{ch} |a_\textnormal{p}(t')|^2 \right)},
\label{eqn:chirping}
\end{equation}
where $C_\textrm{ch}$ is the coefficient mapping the amplitude of the pulse to a corresponding ac Stark shift value. Here we have ignored the global phase coefficient.

\begin{figure}[t]
\begin{center}
  \includegraphics[width=120mm]{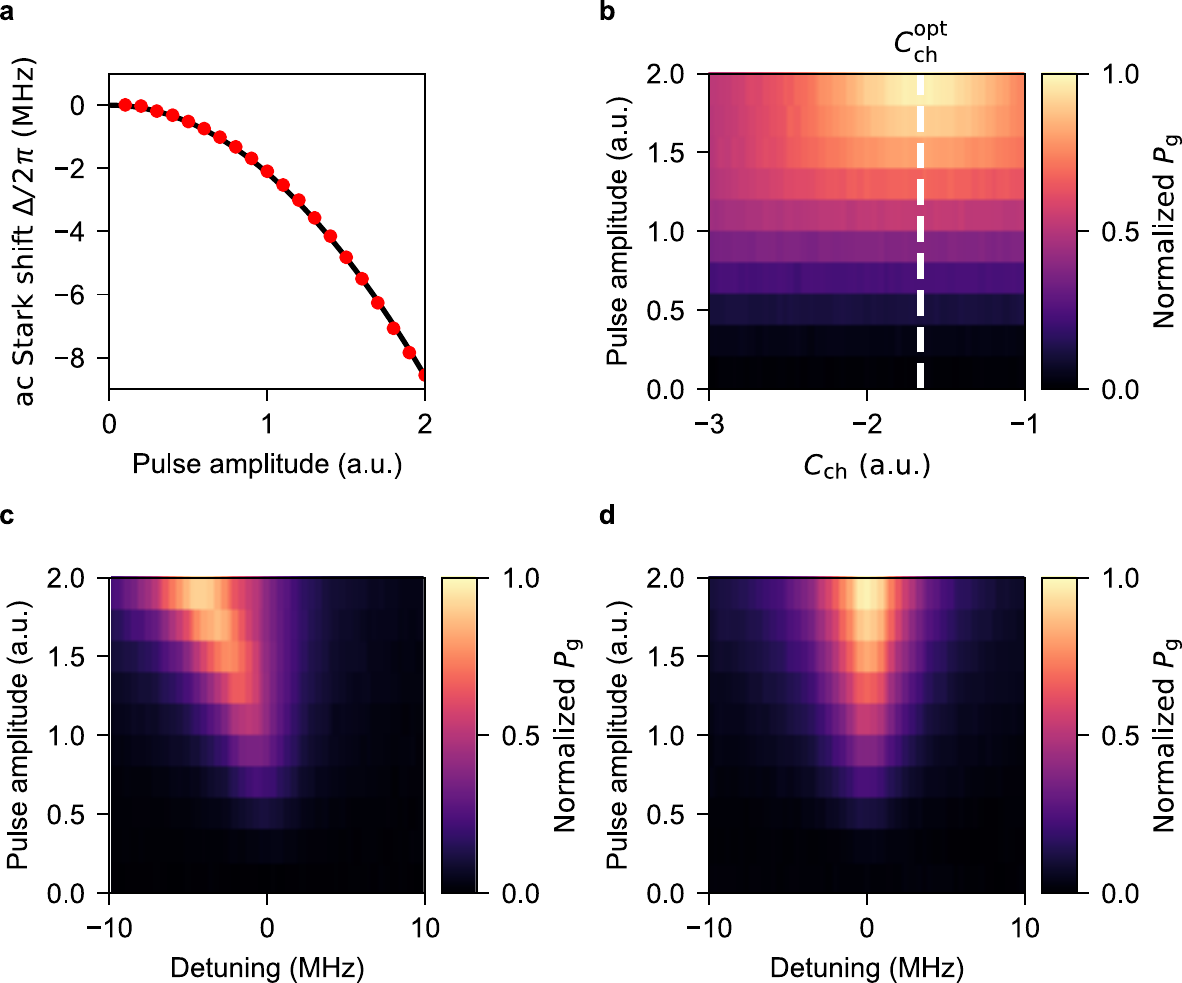} 
\caption{
\textbf{Chirping of the coupling pulse.} (a)~Measured Stark shift $\Delta_\textnormal{f0g1}$ for the $|\textnormal{f}0\rangle$--$|\textnormal{g}1\rangle$ transition as a function of the drive-pulse amplitude (red points). A quadratic fit $C_\textrm{ch} |a_\textnormal{p}(t')|^2$ with coefficient $C_\textrm{ch} = -2.15$ is shown with the black curve.
(b)~Normalized population of qubit $\ket{\textrm{g}}$ state measured after $|\textnormal{f}0\rangle$--$|\textnormal{g}1\rangle$ transition as a function of the chirping coefficient $C_\textrm{ch}$. The optimal parameter $C_\textrm{ch}^\textrm{opt}$ for cancelling the Stark shift caused by the effective coupling pulse is marked by the white dashed line at $C_\textrm{ch}^\textrm{opt} = -1.66$. (c,d)~Measured $\ket{\textrm{g}}$ state population as a function of the maximum effective coupling pulse amplitude and detuning for a coupling pulse (c)~without ac-Stark-shift cancellation and (d)~with cancellation.
} 
  \label{fig:s4}
\end{center}
\end{figure}

In the experiment, the chirped form of the pulse is calculated by first measuring the ac Stark shift of the qubit as a function of the drive amplitude of an effective coupling pulse with a rectangular shape. We show the results of these measurements together with a quadratic fit in Fig.~\ref{fig:s4}(a) with coefficient $C_\textrm{ch} = -2.15$. The maximum drive-pulse amplitude corresponds to an effective coupling rate $g_\textrm{eff}^\textrm{max}/2\pi = \SI{2.2}{\mega\hertz}$. We measure the qubit population transfer in the $\ket{\textrm{f0}}$--$\ket{\textrm{g1}}$ transition as a function of coupling pulse drive frequency and find the optimal frequency where the transfer is maximal for each drive amplitude. The shift of this frequency relative to the optimal frequency at low drive amplitude corresponds to the ac Stark shift. We calculate the necessary phase shift at each amplitude in the pulse with Eq.~\eqref{eqn:chirping} with the coefficient obtained from the quadratic fit. Due to effect of filtering on different pulse shapes and non-linearity of the effective coupling at high pulse amplitudes, the fit values are not necessarily optimal and we need to sweep the coefficient to minimize the effect by the Stark shift. 

We measure the $\ket{\textrm{g}}$ state population $P_\textrm{g}$ after excitation transfer from $\ket{\textrm{f}0}$ to $\ket{\textrm{g}1}$ as a function of the chirping coefficient for a cosine coupling pulse in Fig.~\ref{fig:s4}(b). The optimal chirping parameter value $C_\textrm{ch}^\textrm{opt} = -1.66$ matches the region where the $\ket{\textrm{g}}$ state population is the largest, indicating highest transfer between the two states. The difference between the optimal value and the fit value is due to leakage of the cosine coupling pulse power to the image sideband, which limits the effective coupling rate to $g_\textrm{eff}^\textrm{cos}/2\pi = \SI{1.3}{\mega\hertz}$. We show the difference in state transfer depending on the amount of Stark shift for an effective coupling pulse without any Stark-shift cancellation~[Fig.~\ref{fig:s4}(c)] and with the optimized Stark-shift correction~[Fig.~\ref{fig:s4}(d)]. The optimization causes the transfer to become more uniform across all frequencies and drive amplitudes.

\section*{S5. Calibration of the Josephson parametric amplifier for the amplification of single-photon signal}

\begin{figure}[tp]
\begin{center}
  \includegraphics[width=160mm]{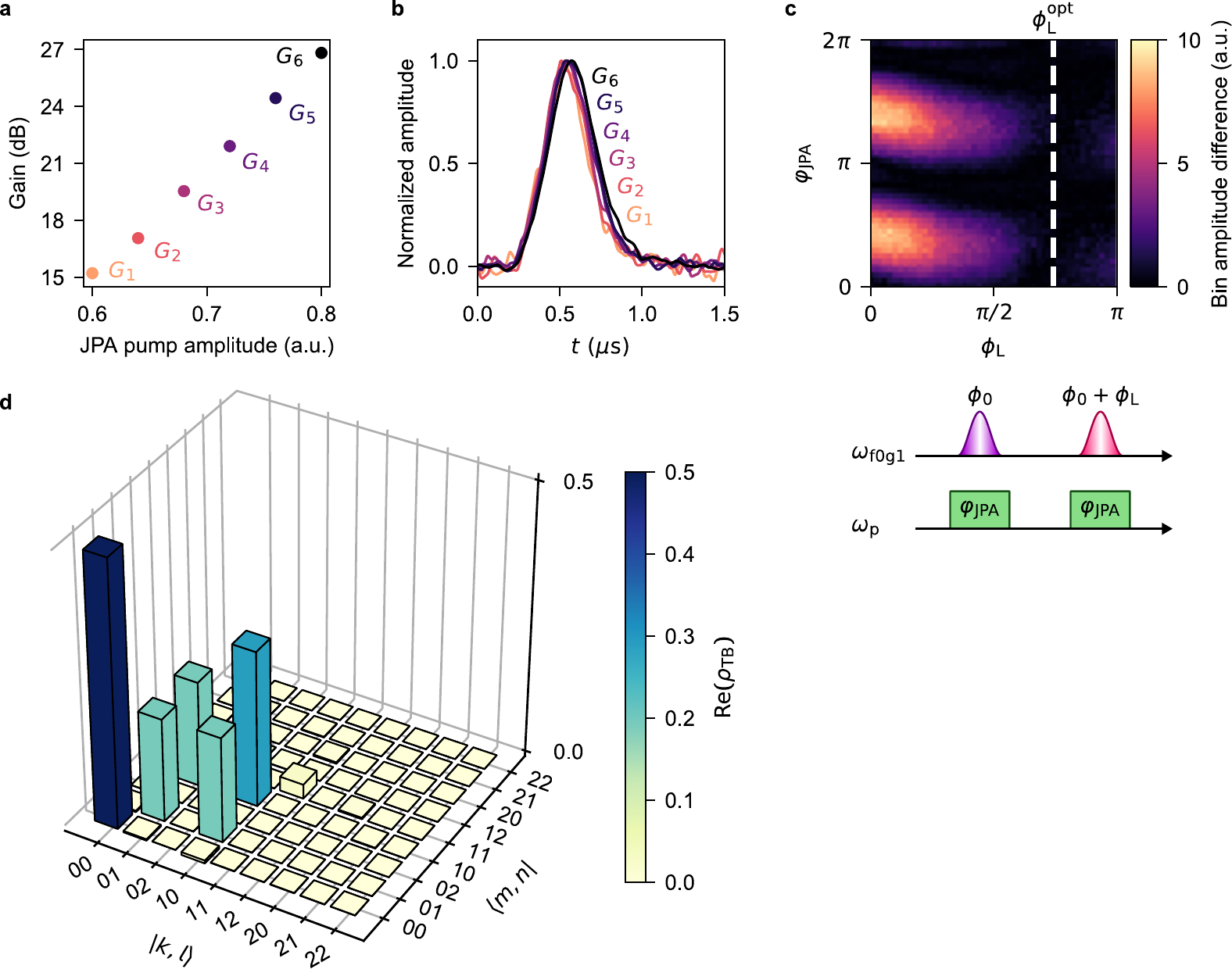} 
\caption{
\textbf{Optimization of time-bin qubit generation and tomography.} (a)~Measured JPA gain as a function of JPA pump amplitude.
(b)~Measured amplified single-photon packet shape as a function of the JPA pump pulse amplitude.
(c)~Absolute difference in measured time-bin qubit signal amplitude between the two bins as a function of JPA pump phase $\varphi_\textrm{JPA}$ and second coupling pulse phase offset $\phi_\textrm{L}$. By modifying the second coupling pulse phase to where the difference is the smallest, $\phi_\textrm{L}^\textrm{opt} = 0.74\pi$, we can correct the offset in phase between the two bins caused by ac Stark shift.
(d)~Measured real part of density matrix elements for a time-bin qubit state $(1/\sqrt{2})(\ket{\textrm{E}} + \ket{\textrm{L}})$. The basis represents the number of photons measured in each temporal mode.
} 
  \label{fig:s5}
\end{center}
\end{figure}

We perform quantum state tomography on the quantum state of a propagating microwave mode with a Josephson parametric amplifier operated in the degenerate mode by driving it with a pulse of phase $\varphi$ to measure a projected quadrature $q_{\varphi}$ of the single photon signal. The amplification anti-squeezes the single photon wave packet along a direction orthogonal to $\varphi$. For the quadrature amplification in tomography, we use a JPA gain of \SI{26.8}{\decibel}~[Fig.~\ref{fig:s5}(a))]. We optimize the amplitude of the JPA pump pulse so that there is no significant distortion in the amplified shape of the single-photon wave packet, as shown in Fig.~\ref{fig:s5}(b).

\section*{S6. Optimization of the coupling pulse for time-bin qubit generation and loss-correction of the time-bin qubit density matrix}

The chirping of the coupling pulse accounts for the ac Stark shift affecting the qubit $|\textrm{f}\rangle$ state, however shift of the $|\textrm{e}\rangle$ state is not corrected by the chirping. During the generation of any coherent superposition state of the time-bin qubit, there is population in the $|\textrm{e}\rangle$ state during the emission of the first bin. The first coupling pulse therefore causes a shift of the $|\textrm{e}\rangle$ state that results in a constant shift of the time-bin qubit phase. To correct this offset and generate the target state, we apply a phase offset to the second coupling pulse relative to the first pulse. To calibrate this phase offset, we measure the difference in amplitude between the signal in the two bins for a coherent time-bin qubit state as a function of the JPA pump phase $\varphi_\textrm{JPA}$ and coupling pulse phase offset $\phi_\textrm{L}$ in Fig.~\ref{fig:s5}(c). At $\phi_\textrm{L}^\textrm{opt} = 0.74\pi$ the difference between the signal amplitude in the two bins is the smallest, corresponding to correction of the phase shift due to ac Stark shift.

For the tomography of the time-bin qubit state, we apply a loss correction to the density matrix in post-processing by reducing the measured complete time-bin qubit density matrix shown in Fig.~\ref{fig:s5}(d) to the single-photon subspace, which corresponds to the time-bin subspace. We pick the submatrix inside the raw data density matrix corresponding to the elements with a single photon in one of the temporal modes. The resulting density matrix is subsequently normalized. This operation can be realized in real-time if one has a single-photon detector or up to high degree with a parity measurement in quantum-non-demolition manner~\citep{bib:kono, bib:besse2018single}.

\section*{S7. Errors in the photonic qubit generation}

In the generation sequence of the photonic qubits, we measure the superconducting qubit state before and after the generation. Here, we discuss the measurement events corresponding to errors in the generation. In 12.2\% of measurement events for the single-rail and in 14.4\% of the time-bin qubit, we measure an initial excited state and a final state corresponding to the ground state. These events correspond to measurements where an initially (thermally) excited qubit has a state prepared and transferred to the photonic qubit. Events corresponding to an initial measurement in the ground state and a final measurement in the excited state, caused mostly by errors in the system transitions, correspond to 2.90\% of the events in the single-rail measurement and to 4.25\% of the events in the time-bin measurement. Some of these errors also occur due to the limited assignment fidelity of our readout. In the case of the single-rail qubit measurements, these events show a relatively large two-photon component in the reconstructed density matrices compared to the other measurement events, possibly due to JPA back-action~\citep{bib:measeffkindel}.

\bibliography{bibnew}